# Magnetic and Transport Studies on Electron-doped $CeFeAsO_{1-x}F_x$ Superconductor

Shen V. Chong, Takahiro Mochiji, Shun Sato and Kazuo Kadowaki

*Institute of Materials Science and Graduate School of Pure & Applied Sciences, University of Tsukuba, 1-1-1, Tennodai, Tsukuba, Ibaraki 305-8573, Japan*

The magnetic and transport behaviors of cerium substituted iron oxy-arsenide superconductor with x = 0.1 to 0.4 fluoride (F) doping have been investigated in this report. Temperature dependent susceptibility and resistivity measurements showed the 0.1 F-doped sample ($CeFeAsO_{0.9}F_{0.1}$) has a superconducting transition temperature ($T_c$) of around 30 K. With increasing doping beyond x = 0.2 $T_c$ saturates to around 40 K. Temperature dependent susceptibility measured in different magnetic fields for the under-doped sample showed Meissner effect in low field and the diamagnetism is still visible up to 1 Tesla, with an obvious magnetic transition below 5 K, perhaps originating from magnetic ordering of the rare earth cerium. The corresponding field dependent resistance versus temperature measurements indicated a broadening of less than 3 K for $T_c$ at mid-point by increasing the field to 5 Tesla indicating rather low anisotropy. An estimated upper critical field of more than 48 Tesla and accordingly an estimated maximum coherence length of 26 Å were obtained confirming the high upper critical field with a short coherence length for this superconductor.

## 1. Introduction

Electron-doped cerium iron oxypnictide superconductor possesses many similar features compared with LaFeAsO superconductor despite the fact that the rare earth cerium carries a magnetic moment while lanthanum does not in the respective oxypnictides. In both superconductors, antiferromagnetic ordering due to spin-density-wave (SDW) instability were observed below 160 K with doping dependent transition temperature behaviors that tread opposite to the doping level.[1,2] Furthermore, there is also a lattice distortion that accompanies each of these magnetic transitions from tetragonal to orthorhombic.[3] The superconducting transition temperature of $CeFeAsO_{1-x}F_x$ saturates at around 41 K at ambient pressure,[2] however in contrary to $LaFeAsO_{1-x}F_x$ a negative pressure dependent $T_c$ was observed from high-pressure resistivity study on optimum doped $CeFeAsO_{0.88}F_{0.12}$ sample.[4] Other spectroscopic studies such as Raman scattering on $CeFeAsO_{1-x}F_x$ at room temperature reveals features associated with complex processes either due to a combination of multi-phonon processes, magnetic fluctuations and/or interband transitions.[5]

In this report, we focus on the magnetic and transport properties of $CeFeAsO_{1-x}F_x$ at different doping level from x = 0.1 to 0.4 with an emphasis on the under-doped Cerium oxypnictide. Temperature dependent mass susceptibility measurements show $T_c$ saturating beyond x > 0.1 at around 40 K. Hall effect measurements confirm this oxypnictide is populated with electron carriers with negative Hall coefficient. The resistance as a function of temperature measurements at different applied magnetic fields support the low anisotropic nature of this class of superconductors with a high upper critical field tolerance.

## 2. Experimental Details

Polycrystalline $CeFeAsO_{1-x}F_x$ oxypnictides were prepared in evacuated sealed quartz tubes by reacting stoichiometric amount of CeAs, Fe, As (or FeAs), $Fe_2O_3$, and $CeF_3$. CeAs was prepared by reacting metallic cerium and arsenic chips at 500 °C for 15 h, 850 °C for 5 h and finally at 900 °C for 5 h. FeAs was prepared by reacting pressed pellets of finely ground iron and arsenic at 700 °C for 10 h. Both metallic arsenides were also synthesized in evacuated sealed quartz tubes. The stoichiometric mixture of the cerium oxypnictide components was thoroughly mixed and pressed into pellets prior to heat treatment. In order to ensure optimum reaction of the reactants, the pellets were reacted sequentially first at 500 °C for 15 h then 850 °C for 5 h followed by 1000 °C for 24 h before finally at 1200 or 1250 °C for 45 h or more.

The crystal structure of the resulting compound was examined by powder x-ray diffraction (XRD) using Cu Kα radiation at room temperature. Electrical resistivity was measured by a standard 4-probe technique down to 4 K. Magnetization measurements were carried out in a Quantum Design SQUID magnetometer down to 2 K. Hall coefficient and magnetic field dependent resistivity measurements were conducted via a Quantum Design physical property measurement system (PPMS) down to 4 K in magnetic fields up to 5 Tesla. The microstructure of the samples was investigated by scanning electron microscope (SEM) operated at 20 kV electron beam energy.

## 3. Results and Discussion

Powder XRD pattern of the undoped CeFeAsO sample displays only Bragg reflections originating from the rare-earth (Re) iron oxy-arsenide (ReFeAsO) tetragonal crystal structure.[2)] With increasing F-doping, it was noted that impurities begin to emerge with Ce-oxides being the main components as indicated by arrows in Fig. 1a. Fig. 1b displays the field-cooled (FC) magnetization versus temperature (*M-T*) curves from the different x values of fluoride doping. Meissner effect was clearly observed in all the fluoride doped samples while that of the undoped sample only display Curie-Weiss paramagnetic behavior above 5 K with a small magnetic transition at very low temperature (~4.02 K), which likely origin was from the ordering of the rare-earth cerium magnetic moment in the quaternary compound. The onset $T_c$ for the x = 0.1 sample is around 30 K which increases to ~40 K for x = 0.2 and beyond, while the cell volume calculated from XRD decreases with increasing doping level. An estimated superconducting volume fraction of about 25 % was obtained for the x = 0.2 doped sample at 5 K under 50 Gauss applied field. The presence of greater amount of impurities clearly contributes to the magnetization curves for the higher doped samples with those of x = 0.3 and 0.4 showing an upturn in magnetization below 100 K, perhaps due to ferromagnetic transition of iron containing compounds.

Fig. 2a shows the *M-T* behavior of $CeFeAsO_{0.9}F_{0.1}$ superconductor in the presence of different applied magnetic field strength. A magnetic transition similar to that observed in the undoped sample below $T_c$ was also seen here, being most obvious at 0.1 Tesla (T) applied field. Fig. 2b shows with increasing field strength, this magnetic transition temperature shifts to lower temperature from 3.0 K at 0.1 T to 2.5 K at 1 T. A similar trend was also observed in $GdFeAsO_{1-x}F_x$ superconductor,[6)] but in the latter case a larger shift in transition temperature was observed (8.1 K at 0.1 T to 4.0 K at 1 T) as consistent to the fact that Gd carries a larger magnetic moment. This further confirms the magnetic transition below $T_c$ in Fig. 2 originates from the rare-earth cerium and supports the co-existence of superconductivity and magnetic order in this class of superconductor. The high temperature region of the *M-T* curves in Fig. 2a follows the Curie-Weiss-like behavior, $1/(\chi_T - \chi_0) = (T + \theta)/C$, with the fitted data giving Curie-Weiss temperatures $\theta < 0$ (ca. -70.3 K from 1 T plot) observed in all the *M-T* curves examined from 0.005 to 1 T applied field. This strongly supports a predominant antiferromagnetic ordering in this 0.1 fluoride doped Ce-based oxypnictide.

Temperature dependent resistance measurements in different magnetic fields show that the onset superconducting transition temperature decreases very slowly with increasing magnetic field strength (Fig. 3). Fig. 3b shows that $T_c$ at zero resistance broadened by 7.0 K from 0 to 5

T while that of $T_c$ at mid-point (between 90 and 10 % transition) was only by 2.8 K. The linear slope near $T_c$ in the field dependent $T_c$ at mid-point data was used to estimate the upper critical field, $H_{c2}$, by the Werthamer-Helfand-Hohenberg equation: $H_{c2} = -0.693(dH_{c2}/dT)T_c$.[7] With $(dH_{c2}/dT) = 2.3$ T/K, an upper critical field of 48.8 T was obtained based on a $T_c$ of 30 K. Since this upper critical field represents the lowest limit of the anisotropic polycrystalline superconductor, the maximum coherence length, thus, can be estimated to be ~26 Å. These results together with those obtained from other rare-earth oxypnictides show the low anisotropic nature of this class of superconductor with a high upper critical field.[8-10] All these point toward potentials for practical applications for these superconductors.

Hall effect measurements show a linear relationship for the transverse Hall voltage ($V_y$) as a function of magnetic field in the range of ± 5 T measured at different temperatures. The slope, $dV_y/dH$, remains negative for all temperatures above $T_c$, indicating the predominant charge carriers are electrons. The Hall coefficient, $R_H$, computed from these linear slopes, displays a slight temperature dependent trend which were also observed in other rare-earth oxypnictides.[8] Taking $R_H$ at 100 K, the charge carrier density ($n = 1/e \cdot R_H$) was found to be ~$8.3 \times 10^{20}$ cm$^{-3}$ in good agreement with the ideal carrier density of $6.73 \times 10^{20}$ cm$^{-3}$ if every fluoride donates an electron into the system in $CeFeAsO_{0.9}F_{0.1}$. The charge mobility ($\mu$) at 100 K was found to be around 339 cm$^2$/V/s.

SEM imaging of the $CeFeAsO_{0.9}F_{0.1}$ superconductor (Fig. 5a) shows a densely packed bulk sample surface containing voids in some areas and grains having plate-like morphology which strongly supports the layered nature of these superconductors. The average lateral grain size was found to be around 15 μm. Magnetization hysteresis measurements at 5 K (Fig. 5b) indicate a strong ferromagnetic background owing to the presence of Fe-based impurities such as FeAs and Fe-oxides. Using the extended Bean model for intragrain critical current density estimation ($J_c = 30\Delta M/d$, where $\Delta M$ is the difference in magnetization ($M_+ - M_-$) in Gauss and $d$ is the average grain size),[11] the calculated $J_c$ was found to be only slightly field dependent above 0.5 T with an average $J_c$ of around $5.5 \times 10^5$ A/cm$^2$. At low magnetic field a maximum $J_c$ of $1.5 \times 10^6$ A/cm$^2$ was obtained. These values are in good agreement with those from the other rare-earth oxy-arsenides.[10,12]

In summary, cerium-based iron oxy-arsenides with $T_c$ equals 30 K and above were prepared in this study. Powder XRD shows the cell volume decreases with increasing doping level with $T_c$ plateauing at 0.2 F-doping and beyond at 40 K. It was also discovered that the amount of impurities also increases at higher doping level. Hall effect measurements confirm the electron doping nature of this superconductor with a slight temperature dependent Hall coefficient being observed. Temperature dependent resistance measurements in fields up to 5 T indicate the low anisotropy nature of this material with a high upper critical field, $H_{c2} > 48$ T and the coherence length of 26 Å. A maximum estimated intragrain critical current density of ~1.5 MA/cm$^2$ was found in the under-doped $CeFeAsO_{1-x}F_x$ superconductor.

**Acknowledgments**

The authors would like to thank T. Goya, K. Yamaki, S. Hashimoto and T. Yamamoto for their respective assistance in sample preparation, physical properties measurements, and fruitful discussion. SVC would also like to thank the Japan Society for the Promotion of Science (JSPS) for their financial assistance.

# FIGURES

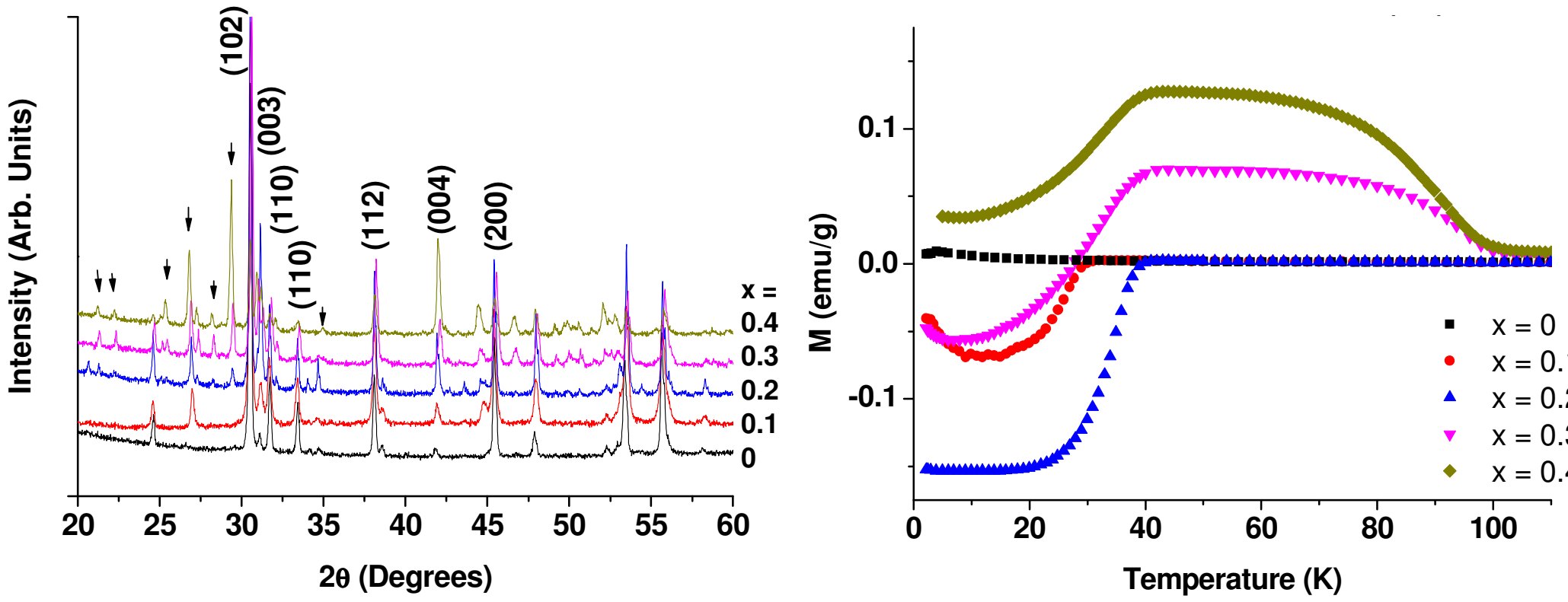

(a) (b)

Fig. 1. Power XRD patterns of $CeFeAsO_{1-x}F_x$ oxypnictides (a) and the corresponding FC temperature dependent magnetization measurements (b). The arrows in (a) indicate some of the impurity phases in the samples.

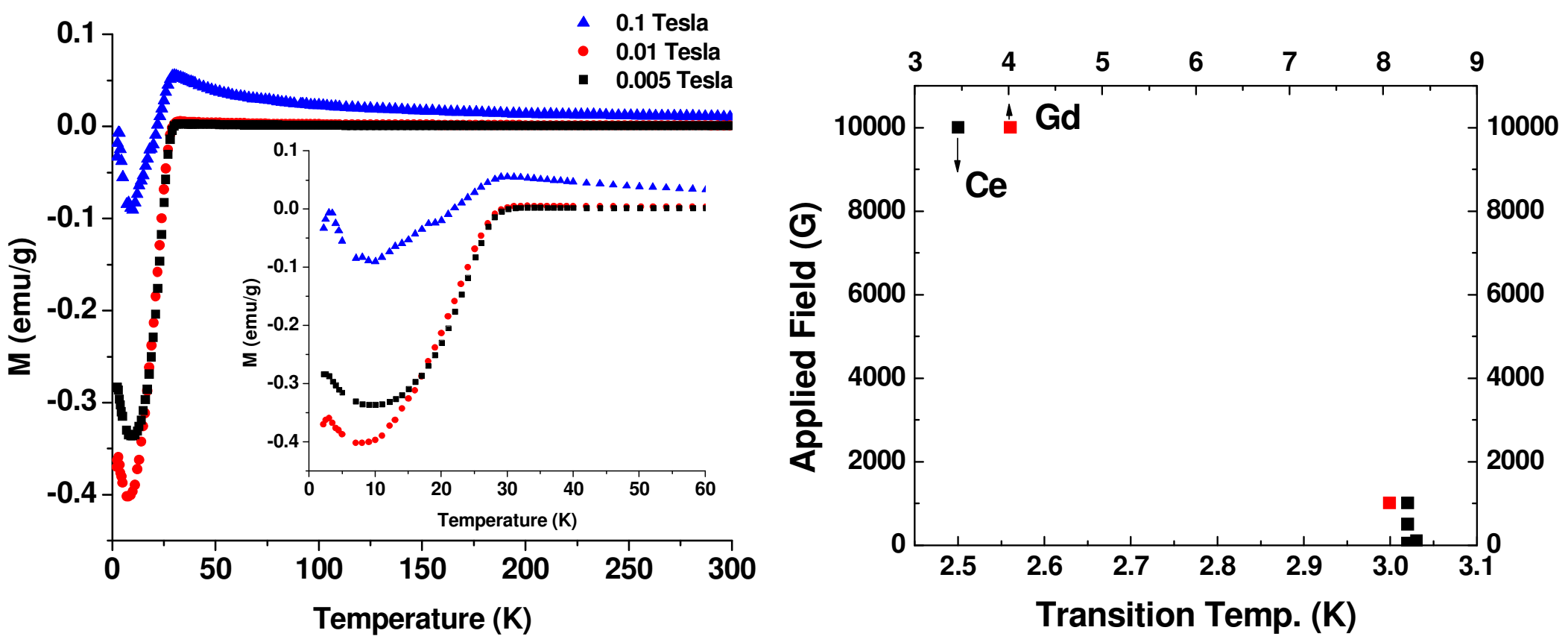

(a) (b)

Fig. 2. *M-T* of $CeFeAsO_{0.9}F_{0.1}$ measured in different magnetic fields (a); inset is the enlarged low temperature part of the plot showing a magnetic transition due to Ce below $T_c$ . (b) shows a comparison between the Ce and Gd low temperature magnetic transition with increasing field from the respective iron oxy-arsenide superconductor.

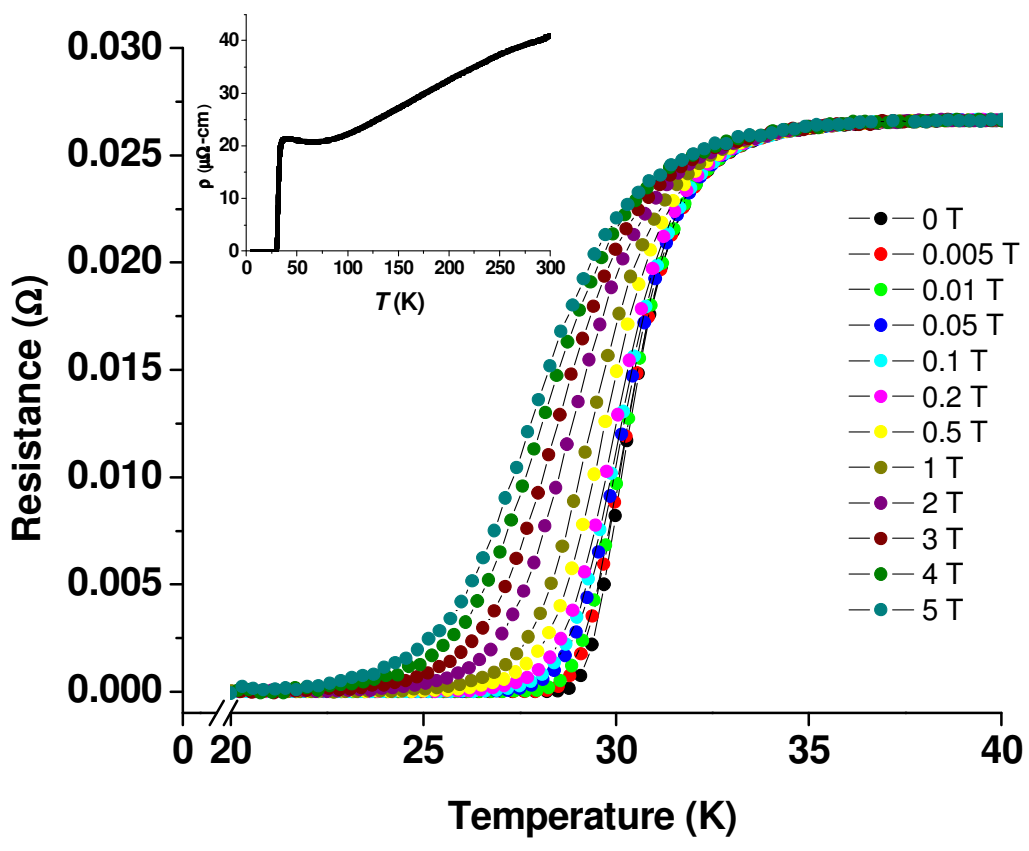

(a)

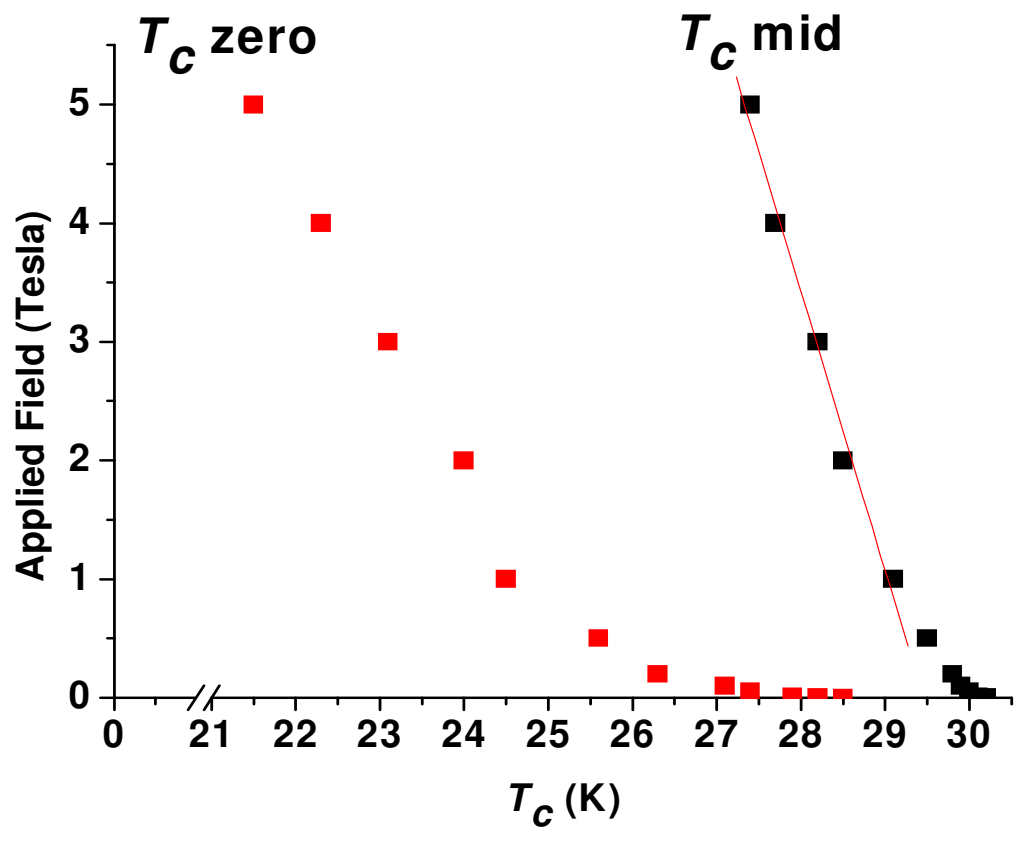

(b)

Fig. 3. Temperature dependent resistance measured in different magnetic fields for $CeFeAsO_{0.9}F_{0.1}$ (a) with the inset showing the resistivity measured at zero applied field from a different sample giving agreeable $T_c$. (b) shows the behavior of $T_c$ at zero and mid-point resistance with increasing magnetic field strength.

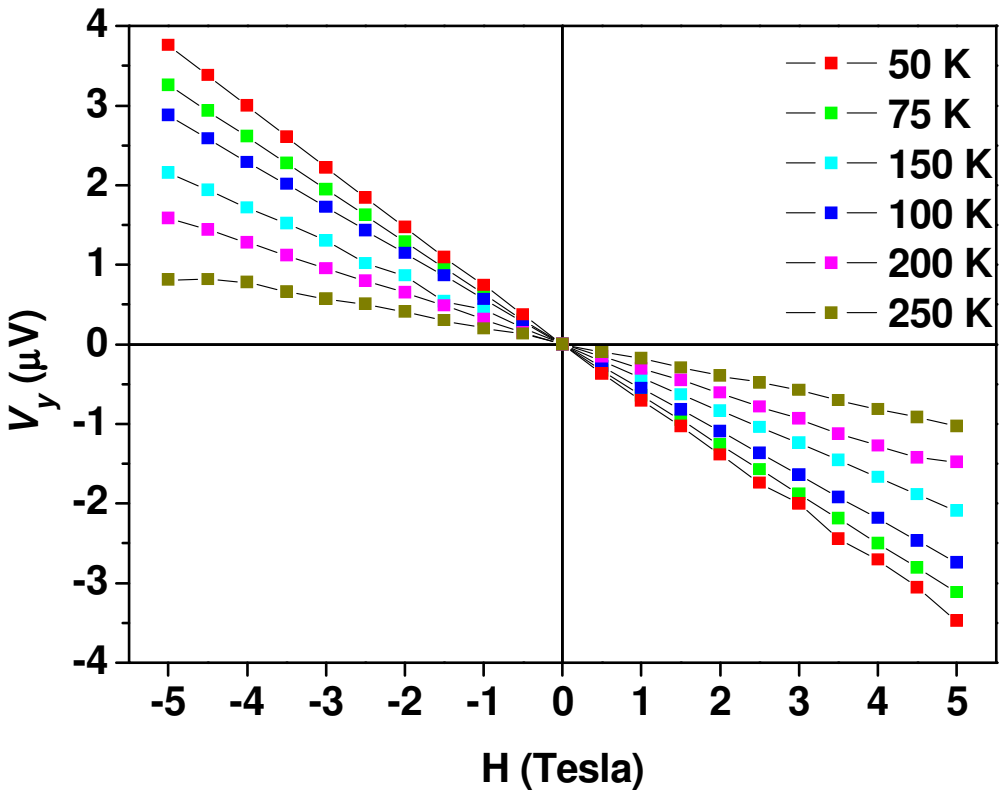

(a)

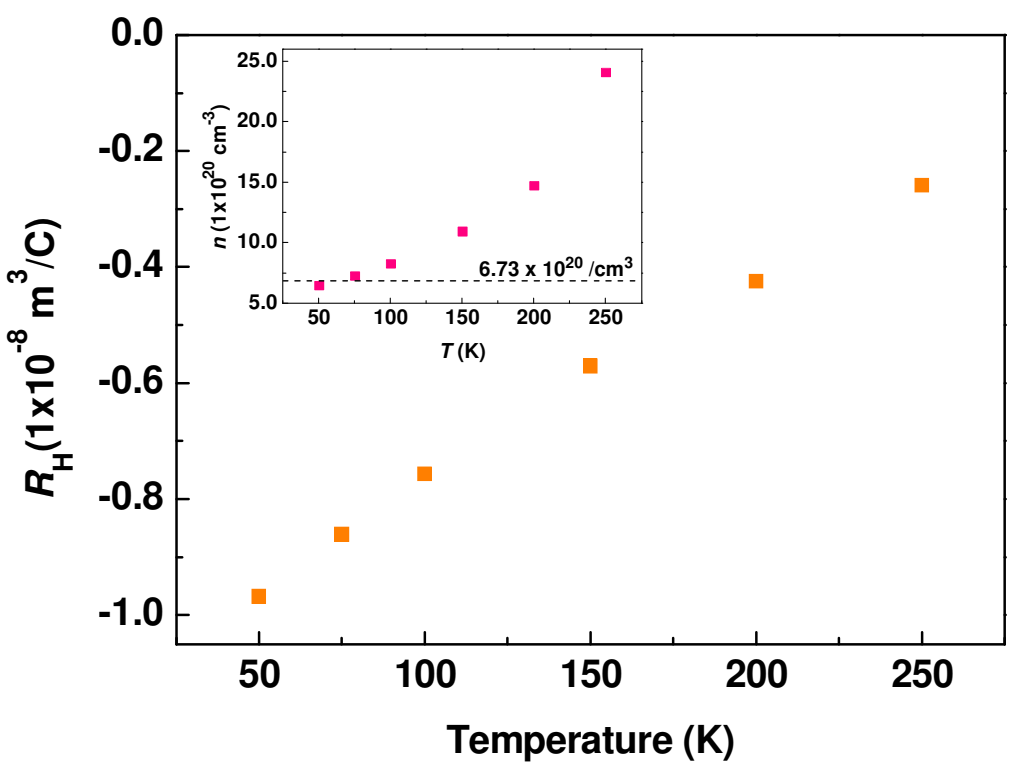

(b)

Fig. 4. Hall effect measurements showing the transverse Hall voltage measured from -5 to +5 T (a), and the resulting Hall coefficient as a function of temperature (b). The inset in (b) shows the corresponding charge carrier density (*n*).

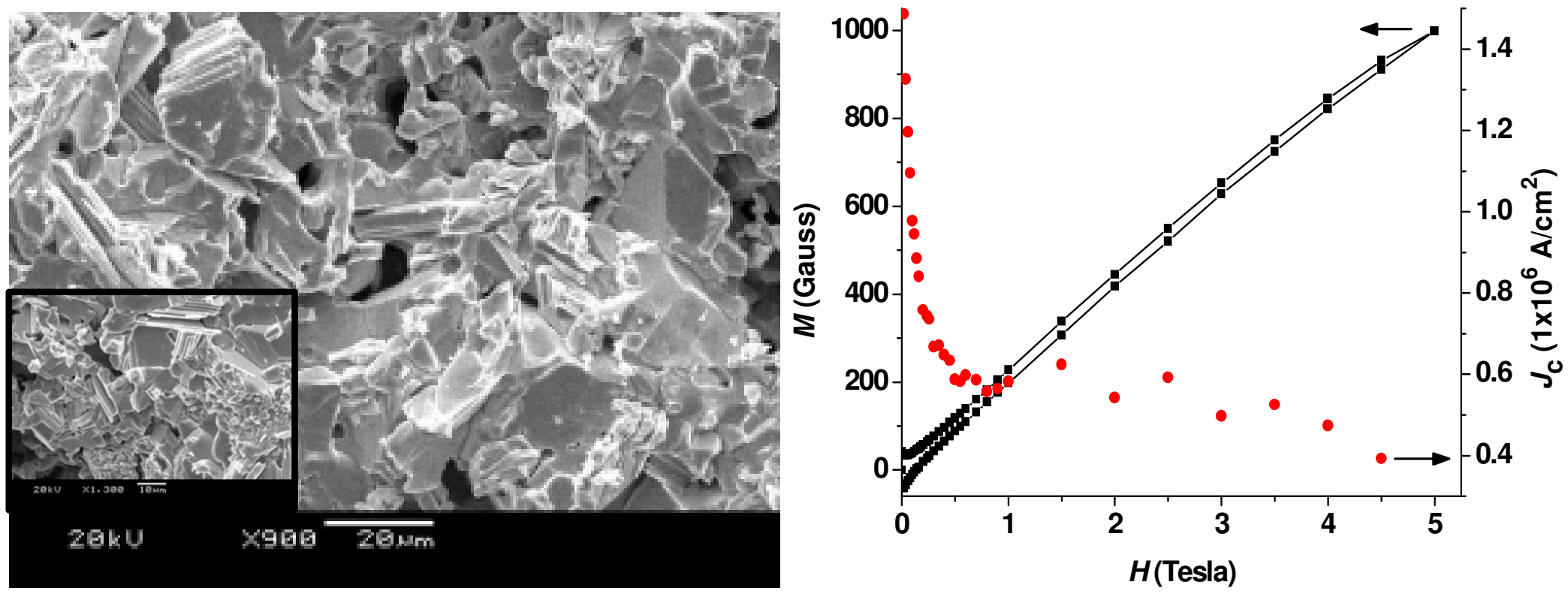

(a) (b)

Fig. 5. SEM of $CeFeAsO_{0.9}F_{0.1}$ showing plate-like morphology of some grains (a) and *M*-*H* measurement on the same sample up to 5 T (b). The other dotted data in (b) shows the estimated intragrain critical current density obtained from the modified Bean Model.